\documentclass{article}
\usepackage[a4paper, margin=1in]{geometry}  % Optional: adjust margins
\usepackage{authblk}  % For affiliations
\usepackage{amssymb}
\usepackage{natbib}
\usepackage{graphicx}
\usepackage{bm}
\usepackage{amsmath}  % Minimum required
\usepackage{hyperref}
\newcommand\pd[2]{\frac{\partial #1}{\partial #2}}

\newcommand{\mathsfbi}[1]{\bm{\mathsf{#1}}}
\title{Optimal airfoils in the intermediate Reynolds number range}

\author[1]{Gleb Zhdanko}
\author[1]{Dmitry Kolomenskiy}

\affil[1]{Center for Materials Technologies, Skolkovo Institute of Science and Technology, Moscow, Russia}

\date{\today}  % Or \date{2026} for fixed date

\begin{document}

\maketitle

\begin{abstract}
    We revisit a classical airfoil design problem: the search for shapes that maximize aerodynamic performance metrics, targeting the underexplored intermediate Reynolds-number regime between 1 and 3000, relevant to small animals and miniature vehicles. The problem is formally stated as the glide ratio or the endurance factor maximization for Joukowski airfoil profiles and for more general airfoil shapes with adjustable position of the maximum camber, under steady inflow. It is solved numerically by a hybrid approach combining stochastic search and direct parameter sweep, and using a steady laminar Navier--Stokes solver based on conformal mapping and second-order finite-difference discretization.
    Zero-thickness cambered airfoils are found to be globally optimal, within the Joukowski family, across the entire Reynolds-number range considered.
    The optimal angle of attack decreases monotonically with $Re$, whereas the optimal camber varies non-monotonically,
    reaching a pronounced maximum near $Re \approx 40-50$ before declining at higher $Re$.
    At low Reynolds numbers ($Re \lesssim 100$), a broad family of cambered shapes performs within a few per cent of the optimum, indicating weak sensitivity to geometrical parameters.
    In contrast, for $Re \gtrsim 1000$, the performance landscape becomes sharply localized around a single preferred design, for which geometric refinement is critical.
\end{abstract}
\section{Introduction}
\label{sec:introduction}
Lift and propulsion mechanisms in fluids have been extensively studied across a wide range of the Reynolds numbers (\(Re\)). At high \(Re\) ($>10^4$),
typical of aeronautical applications, the literature is rich with theoretical, computational, and experimental insights.
Decades of research have established that cambered thin-plate airfoils ensure optimal lift-to-drag ratios for the range of \(Re\) up to $10^5$. Low-\(Re\) regimes (\(Re \ll 1\)),
typical of microbial swimmers, have also received substantial attention.
Here, the Stokes flow approximation enables analytical solutions yielding explicit formulas for the fluid-dynamic forces. Intermediate \(Re\) ($1 \lesssim Re \lesssim 10{,}000$),
prevalent in biological locomotion, have historically received less attention. This gap likely stems from analytical intractability (viscous and inertial effects compete without simplifying assumptions) and experimental challenges: neither traditional high-\(Re\) wind tunnels nor microfluidic setups can reliably cover this range of \(Re\).
Recent technological advances in bio-inspired micro air vehicles have spurred practical interest in this flow regime.

Studies on insect flight (\(Re \sim 10\)--$1000$) often employ flat, cambered, or even corrugated thin foils as wing models.
Yet, systematic efforts to identify optimal airfoil shapes remain scarce, with most work prioritizing kinematics over geometry.
A notable exception is the shape-optimization study by \citet{srinathOptimalAirfoilShapes2009},
which explored a broad parameter space of airfoil contours at fixed angles of attack $(\alpha = 4^\circ$ and $12^\circ )$
using adjoint optimization. However, the complex shapes obtained therein reflect the \(\alpha\) constraint rather than the necessity
for non-zero thickness. Our present study addresses this problem through a low-dimensional parameterization of airfoil profiles,
enabling efficient exploration of the combined effect of main features of the shape (i.e., thickness and camber) and angle of attack across intermediate \(Re\).
By coupling a parametric geometrical description with Navier--Stokes simulations, we identify optimal airfoils within the Joukowski family, then more general airfoil shapes with adjustable position of the maximum camber. We relate the observed trends of the optimal parameters as functions of $Re$ to the competing effect of camber-enhanced
lift and inertia-driven separation.
\section{Mathematical model}
\label{sec:mathemtaical_model}
\subsection{Governing equations}
\begin{figure}[h]
    \centering
    \includegraphics[width=\textwidth]{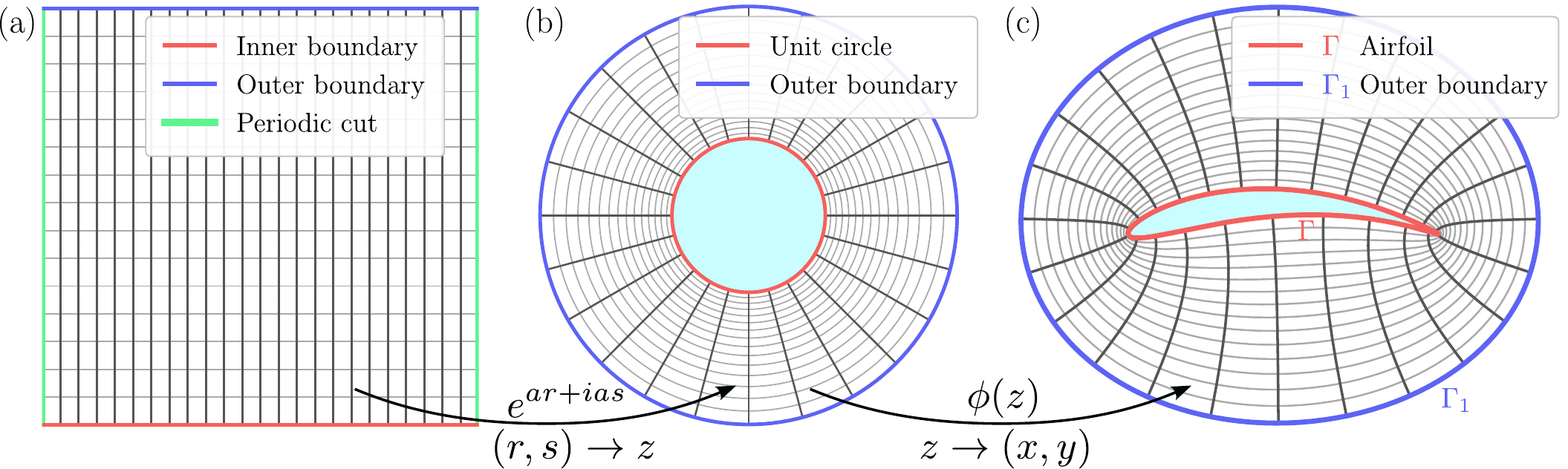}
    \caption{The three-stage conformal mapping: (a) rectangular computational domain in $(r,s)$
        coordinates; (b) intermediate annular domain in the
        $z$-plane; (c) physical domain with the body-fitted grid to the airfoil surface. Inner and outer boundaries are shown in red and blue, respectively.}
    \label{fig:conformal_diagram}
\end{figure}

Airfoil optimization requires flow solutions for many candidate profiles, making repeated mesh generation computationally expensive.
We avoid this by using conformal mapping, which simultaneously defines the airfoil, generates an orthogonal grid, and yields the metric terms for the transformed Navier--Stokes equations.
The map is constructed in two stages, mapping the airfoil exterior to the computational domain via an intermediate annular domain (see Fig~\ref{fig:conformal_diagram}).
The starting point is a
rectangular grid in $(r,s)$ coordinates, where $r$
is the radial coordinate and $s$ the periodic angular coordinate. The first map,
$
    z = e^{ar + \mathrm{i}as},$
deforms this rectangular grid into an O-grid in the complex $z$-plane: lines
of constant $r$ become concentric circles, and lines of constant $s$ become
radial rays. The parameter $a \in \mathbb{R}$ controls the grid stretching.
The second map, $\phi$, from the annulus onto the airfoil
exterior,
$
    x + \mathrm{i}y = \phi(z),
$
deforms the O-grid into a body-fitted mesh, with the inner boundary
conforming to the airfoil surface $\Gamma$ and the outer boundary mapped to the
far field boundary $\Gamma_1$.
We will examine two sets of results using two different maps $\phi(z)$, respectively: Joukowski transform (section \ref{sec:jouk}) and Wegmann transform (section \ref{sec:wegmann}).
With the composite mapping defined, the incompressible steady Navier--Stokes
equations in vorticity-stream function form
can be rewritten in $(r,s)$ coordinates. Denoting $A = \Re(\phi'(z)az)$
and $B = \Im(\phi'(z)az)$, the Jacobian of the transformation
$(r,s)\to (x,y)$ is
$
    \mathsfbi{J}= \begin{pmatrix} A & -B \\ B & A \end{pmatrix},
$
and the governing equations are written in computational coordinates
\begin{eqnarray}
    \label{eq:TransformedNS}
    \frac{1}{\det \mathsfbi{J}}\left(\frac{\partial \psi}{\partial s}\frac{\partial \omega}{\partial r}
    -\frac{\partial \psi}{\partial r}\frac{\partial \omega}{\partial s}\right)
    =
    \frac{1}{Re}\frac{1}{\det \mathsfbi{J}}\nabla^2_{r,s}\omega, \\
    \omega = -\frac{1}{\det \mathsfbi{J}}\nabla^2_{r,s}\psi,\label{eq:poissonVorticity}
\end{eqnarray}
where $Re = \rho U_\infty c / \mu$, $\rho$ is the fluid density,
$U_\infty$ the freestream velocity, $c$ the chord length and $\mu$
the dynamic viscosity.
\subsection{Boundary conditions}
\label{sec:boundary_conditions}
On the airfoil surface $\Gamma$, the no-penetration and no-slip conditions are $\partial\psi/\partial \boldsymbol{n} = 0$ and $\partial\psi/\partial \boldsymbol{\tau} = 0$, where $\boldsymbol{n}$ and $\boldsymbol{\tau}$ denote the unit normal and unit tangent vector to $\Gamma$, respectively.
Since conformal maps preserve orthogonality, the conditions retain their form in $(r,s)$ coordinates.
From the boundary condition $\partial \psi / \partial s = 0$ on $\Gamma$, it follows that $\psi$ is uniform on the airfoil surface.
Since $\psi$ is defined only up to an additive constant, this value may be chosen as zero, reducing the boundary conditions to $\psi = 0$ and $\partial \psi / \partial r = 0$ on $\Gamma$.

At the outer boundary $\Gamma_1$, uniform horizontal freestream conditions are imposed, which in terms of the
streamfunction reads $\partial \psi/\partial y = 1,\ \partial \psi/\partial x = 0$ on $\Gamma_1$.
Applying the chain rule for the conformal transformation yields the outer boundary conditions:
\refstepcounter{equation}
$$
    \psi_r = B, \qquad \psi_s = A \quad \text{on } \Gamma_1.
    \eqno{(\theequation{\mathit{a},\mathit{b}})}\label{eq:boundary_outer_streamfunction}
$$
At the continuous level, both conditions~\eqref{eq:boundary_outer_streamfunction}
are valid and equivalent representations of the imposed far-field velocity.
However, in the discrete formulation, prescribing both streamfunction
derivatives simultaneously leads to an ill-conditioned system when coupled
with the Poisson equation~\eqref{eq:poissonVorticity}. Condition
\eqref{eq:boundary_outer_streamfunction}$a$ is retained as it directly
controls the mass flux through the outer boundary, whereas condition
\eqref{eq:boundary_outer_streamfunction}$b$ is redundant given the
irrotationality of the undisturbed freestream and therefore dropped in favour
of the vorticity condition $\omega = 0$. This substitution restores
well-posedness of the discrete system without altering the physical model,
yielding the system
\begin{equation}
    \label{eq:TransformedNSwithBC}
    \begin{aligned}
        \left(\frac{\partial \psi}{\partial s}\frac{\partial \omega}{\partial r}
        - \frac{\partial \psi}{\partial r}\frac{\partial \omega}{\partial s}\right)
               & = \frac{1}{Re}\nabla^2_{r,s}\omega,              \\
        \omega & = -\frac{1}{\det\mathsfbi{J}}\nabla^2_{r,s}\psi,
    \end{aligned}
    \quad\text{with B.C.}\quad
    \begin{aligned}
        \psi                            & = 0, \quad \frac{\partial\psi}{\partial r} = 0
                                        &                                                & \text{on }\Gamma,   \\
        \frac{\partial\psi}{\partial r} & = B, \quad \omega = 0
                                        &                                                & \text{on }\Gamma_1.
    \end{aligned}
\end{equation}

\subsection{Joukowski conformal map}
\label{sec:jouk}
The Joukowski map is
defined on the exterior of the unit disk ($\mathbb{C}\setminus D_1$) by analytic formula
$\widetilde{\phi(z)} = z + z_0 + k^2/(z + z_0)$,
where three real-valued parameters fully determine the airfoil shape:
$k \geq 0$ controls the thickness, $\mu_x = \mathrm{Re}(z_0)$ controls
the trailing-edge geometry,
%  --- $\mu_x < 0$ produces a cusped trailing 
% edge,
% and $\mu_x > 0$ shifts the cusp toward the leading edge ---
and
$\mu_y = \mathrm{Im}(z_0)$ introduces camber. Despite this low-dimensional parameterization,
Joukowski airfoils represent a wide variety of streamlined shapes. This is an analytical mapping, therefore its numerical evaluation is computationally highly efficient and precise.

Since conformality requires a non-zero derivative throughout the domain,
% Conformal map is holomorphic function from one open subset of complex plane to another with non-zero derivative everywhere in the domain space, therefore 
constraints on $k$ and $z_0$ must be imposed.
% The derivative of the map is
% $$\phi(z)' = 1 - \frac{k^2}{(z+z_0)^2}.$$
For a conformal map from the exterior of the unit disk to the exterior of
the airfoil, two conditions must hold: the shifted centre $z_0$ must lie
inside the unit disk, ensuring that the denominator $z + z_0$ is
non-vanishing in $\mathbb{C}\setminus D_1$, and $\widetilde{\phi'(z)} \neq 0$
throughout $\mathbb{C}\setminus D_1$.

Preliminary computations over the full parameter space, consistent with
\citet{sunadaAirfoilSectionCharacteristics1997}, confirm that airfoil
with a cusped trailing edge ($\mu_x \leq 0$) and upward camber
($\mu_y \geq 0$) achieve higher performance metrics; the parameter
space is therefore restricted to
% Profiles with upward camber and cusp trailing edge satisfies the following constraints: 
\begin{equation}
    \label{eq:parameter_constraints}
    \mu_x \in [-1,0],\quad
    \mu_y \in [0,\sqrt{1-\mu_x^2}\ ),\quad
    k \in [0, \sqrt{1-\mu_y^2} - |\mu_x| \ ).
\end{equation}
The airfoil thickness decreases monotonically as $k$ increases
toward its upper bound; in the limit $k \to \sqrt{1 - \mu_y^2}
    - |\mu_x|$ the airfoil degenerates to a flat plate.

Two further operations complete the map definition: normalization to the unit chord and introduction of the angle of attack $\alpha$.
Defining $c = |z_{x_{\max}} - z_{x_{\min}}|$ as the distance between the
leftmost and rightmost points, which coincides with the chord length for
the thin airfoils considered here, the normalized map is
$
    \widehat{\phi(z)} := \widetilde{\phi(z)} / c.
$
The angle of attack $\alpha$ is then introduced by composing $\widehat{\phi}$ with a
rigid rotation:
$
    \phi(z) := \widehat{\phi(z)}\mathrm{e}^{\mathrm{i}\alpha}.
$

The composite normalized Joukowski map, with $\alpha$ as an additional parameter,
thus takes the form
\begin{equation}
    \phi(z) = \frac{e^{\mathrm{i}\alpha}}{|z_{x_{\max}} - z_{x_{\min}}|}
    \left( z + z_0 + \frac{k^2}{z + z_0} \right), \label{eq:full_map}
\end{equation}
with the parameter constraints given by~\eqref{eq:parameter_constraints}.
% \subsection{Wegmann transform}
% Joukowski mapping provides control over the two most important geometric parameters: thickness and camber.
% However, it does not allow independent variation of the chordwise position of maximum camber $p$,
% which is known to be a third key parameter for thin airfoils. To investigate the effect of $p$ and identify
% its optimal values, we adopt the numerical conformal mapping approach proposed by Wegmann [cite original paper].
% This method consists of two steps. First, the airfoil is transformed into a smooth near-circular contour using the inverse K\'{a}rm\'{a}n--Trefftz mapping.
% Second, Wegmann's algorithm computes the exterior conformal map from the unit circle to this near-circle in the form of a truncated Laurent series.
% Composing this map with the K\'arm\'an--Trefftz transformation yields a conformal map from the exterior of the unit circle directly onto the exterior of the airfoil.
% \label{sec:wegmann}
\subsection{Wegmann transform}
\label{sec:wegmann}
Joukowski mapping provides control over the two most important geometric parameters: thickness and camber.
However, it does not allow independent variation of the chordwise position of maximum camber $p$, which is known to be a third key parameter for thin airfoils.
To investigate the effect of $p$, we generate the target profile from a four-parameter family and compute the corresponding map numerically following \cite{wegmannIterativeMethodConformal1986}.

The airfoil is parametrized similarly to NACA profiles. The mean camber line $y_c(x)$, with maximum camber $f$ at chordwise position $p$ is defined as
\begin{equation}
    y_c(x) =
    \begin{cases}
        \dfrac{f}{p^2}\bigl(2px - x^2\bigr),              & x < p,    \\[2ex]
        \dfrac{f}{(1-p)^2}\bigl(1 - 2p + 2px - x^2\bigr), & x \geq p,
    \end{cases}
    \label{eq:camberline}
\end{equation}
and a symmetric thickness distribution $y_t(x)$ is added normal to it. The thickness follows the NACA four-digit polynomial, with a factor $(1-x)^{q-1}$ controlling trailing-edge closure:
\begin{equation}
    y_t(x) = \frac{t}{0.2}\,(1-x)^{q-1}\bigl(0.2969\sqrt{x} - 0.1260\,x - 0.3516\,x^2 + 0.2843\,x^3 - 0.1015\,x^4\bigr).
    \label{eq:thickness}
\end{equation}
The exponent $q$ controls the trailing-edge shape: $q=1.5$ gives a Joukowski-like cusp, $q=1$ the wedge of the standard NACA series. We fix $t = 0.01$ and $q = 1.5$ throughout, leaving $(f, p, \alpha)$ as free parameters. The resulting profiles have the leading-edge radius of $\approx 0.011\%$ of the chord length and the maximum thickness near the leading edge of $\approx 1.2\%$ of the chord length.
The map is then computed in two steps. First, the inverse K\'arm\'an--Trefftz map is computed,
\begin{equation}
    \frac{z-1}{z+1} = \left(\frac{w-1}{w+1}\right)^{1/n}, \qquad n = 2 - \frac{\tau}{\pi},
    \label{eq:kt}
\end{equation}
with $\tau$ defining the trailing-edge wedge angle, which the algorithm measures from the surface tangents at the trailing edge of each profile. For the cusped trailing edges considered here ($q = 1.5$), $\tau$ is small and the exponent is close to $2$. This map removes the trailing-edge corner and deforms the airfoil $\varGamma$ into a smooth near-circular contour $\eta(s)$ parametrized by arclength $s \in [0, 2\pi)$. Second, Wegmann's algorithm computes the exterior map $g$ from the unit circle onto this near-circle. The map must project each boundary point $\mathrm{e}^{\mathrm{i}t}$ to a point on the contour,
\begin{equation}
    g(\mathrm{e}^{\mathrm{i}t}) = \eta(S(t)),
    \label{eq:wegmann}
\end{equation}
where the boundary correspondence $S(t)$ is the unknown. Requiring $g$ to be analytic outside the unit circle turns \eqref{eq:wegmann} into a nonlinear equation for $S(t)$, which Wegmann's method solves by iteration. The converged map is represented as a Laurent series, $g(w) = m_0 w + \sum_{k=0}^{\infty} c_k\, w^{-k}$, of which we retain the first $80$ coefficients.
Composing $g$ with the forward K\'arm\'an--Trefftz transformation~\eqref{eq:kt} results in the conformal map from the exterior of the unit circle directly onto the exterior of the airfoil, yielding $\phi(z)$ analogous to the Joukowski map of Section~\ref{sec:jouk}.
\subsection{Numerical method}
The nonlinear system~\eqref{eq:TransformedNSwithBC} is solved by Newton--Raphson iteration with backtracking line search.
Because convergence is sensitive to the initial guess, different initialization procedures are used for low and high Reynolds numbers.
For $Re < 200$, the potential-flow solution is used as the initial guess. For higher $Re$, a continuation strategy is employed over an increasing sequence of Reynolds numbers,
$Re_1 < Re_2 < \cdots < Re_{\text{target}}$,
with each converged solution used for the next step. As an example, to reach $Re = 1000$, the sequence
$200 \to 300 \to 500 \to 750 \to 1000$
is used.

The linearized system is discretized with second-order finite differences
on a uniform $N_r \times N_s = 450 \times 600$ grid with a periodic
boundary condition in the circumferential direction. The outer boundary is placed
at the distance of $114c$ and $78c$ away from the airfoil for $Re < 150$ and $Re \geq 150$, respectively, where
the larger domain compensates for the slower decay of velocity
perturbations in the low-$Re$ regime \citep{khaliliStokesParadoxCreeping2017}.
The conformal stretching concentrates
nodes near the wall: the first off-wall layer thickness ranges from
$1.3 \times 10^{-4}c$ near the trailing-edge cusp to $7.0 \times 10^{-3}c$
along the remainder of the chord. Solver accuracy is confirmed in
Appendix~\ref{app:validation}.

Once the vorticity--streamfunction system is solved, the lift and drag
coefficients are recovered from the surface pressure $P$ and wall shear
stress $\tau_w$. We use the nondimensionalization $\rho_\infty = 1$,
$U_\infty = 1$, $c = 1$, under which the dynamic pressure is $q_\infty =
    1/2,$ dynamic viscosity is $\mu = 1/Re$ and the lift and drag coefficients reduce to $C_L = 2F_L$,
$C_D = 2F_D$, where $F_L$ and $F_D$ are lift and drag forces.

The surface pressure satisfies $\mathrm{d}P/\mathrm{d}s =
    -(1/Re)(\partial\omega/\partial r)|_{\Gamma}$, which is obtained from the
tangential momentum equation at the wall, with gauge condition
$\int_\Gamma P\,\mathrm{d}s = 0$. The wall shear stress is
$\tau_w = \omega\sqrt{\det\boldsymbol{J}}/Re$.
The lift and drag coefficients follow from integrating the pressure
and shear-stress contributions over the airfoil surface, with the
Jacobian components $A$ and $B$ entering through the coordinate
transformation:
\begin{equation}
    C_D = -2\int_\Gamma P A\:\mathrm{d}s
    - \frac{2}{Re}\int_\Gamma \omega B \:\mathrm{d}s, \quad
    C_L = -2\int_\Gamma P B\:\mathrm{d}s
    + \frac{2}{Re}\int_\Gamma \omega A \:\mathrm{d}s.
\end{equation}
\section{Optimization}
We optimize airfoil profiles over $Re \in [1, 3000]$ in two independent sets of numerical experiments to maximize, respectively, two different performance metrics: $C_L/C_D$ (in the literature termed as lift-to-drag ratio, or glide ratio) and $C_L^{3/2}/C_D$ (termed as endurance factor, or power factor).
% They are associated with gliding range and flight endurance, respectively (\cite{muellerAERODYNAMICSSMALLVEHICLES2003}).
These are the two most commonly used objectives when designing airfoils for regimes that typically constitute most of the flight duration, e.g., cruise and loitering flight.
Since direct search over the full four-parameter space $(k, \mu_x,
    \mu_y, \alpha)$, as defined in \eqref{eq:full_map}, is computationally
expensive, we progressively narrow the parameter domain as follows.

We begin by analysing symmetric airfoils: $\mu_y = 0$, $\mu_x = 0$. This reduction
to a two-parameter problem ($k,\alpha$) makes it feasible to construct the
objective function over a structured grid and examine its global behaviour.
The computed objective surfaces reveal that, within the class of symmetric
airfoils, the highest performance is consistently achieved for zero-thickness airfoils,
corresponding to a flat plate. To determine whether this tendency persists when camber and
trailing-edge geometry are free, we proceed to the full four-parameter
space of Joukowski foils, employing three complementary global search strategies in
parallel: Sobol sequence sampling, Covariance Matrix Adaptation Evolution
Strategy (CMA-ES), and Bayesian optimization. All three strategies consistently identify zero-thickness airfoils
with $\mu_x \approx 0$ as optimal across the entire $Re$ range, in agreement
with existing results for low-Reynolds-number airfoils
\citep{sunadaAirfoilSectionCharacteristics1997}.
Accordingly, the parameter $\mu_x$ is set to $0$ and $k$
is set to the zero-thickness
limit for each $\mu_y$, as given by the upper bound in~\eqref{eq:parameter_constraints}. This reduction implies
$
    k = \sqrt{1 - \mu_y^2},\footnote{In practice, $k$ is set to
$\sqrt{1-\mu_y^2} - \varepsilon$ with $\varepsilon = 10^{-7}$ to avoid
    the map singularity at the boundary of the admissible domain; this has
    no measurable effect on the aerodynamic coefficients.}
$
leaving $\mu_y$ as the only geometric parameter of the airfoil. For a thin
Joukowski profile, $\mu_y$ is related to the camber $f$ through the image
of the top of the circle ($z = \mathrm{i}$):
${f = \Im\!\left(
        \mathrm{i} + \mu_y\mathrm{i} + (1 - \mu_y^2)/(\mathrm{i} + \mu_y\mathrm{i})
        \right)/c = \mu_y/(2k)},
$
which inverts to give
$
    \mu_y = 2f/\sqrt{1 + 4f^2}.
$
The optimization thus reduces to a two-parameter problem in $(f, \alpha)$ space.
A structured high-resolution grid search is performed over this
plane with steps $\Delta\alpha = 0.5^\circ$ and $\Delta f = 0.5\%$, within
windows centred on the regions identified by the global search.
\section{Results and discussion}
\begin{figure}[ht]
    \centering
    \includegraphics[width=\textwidth]{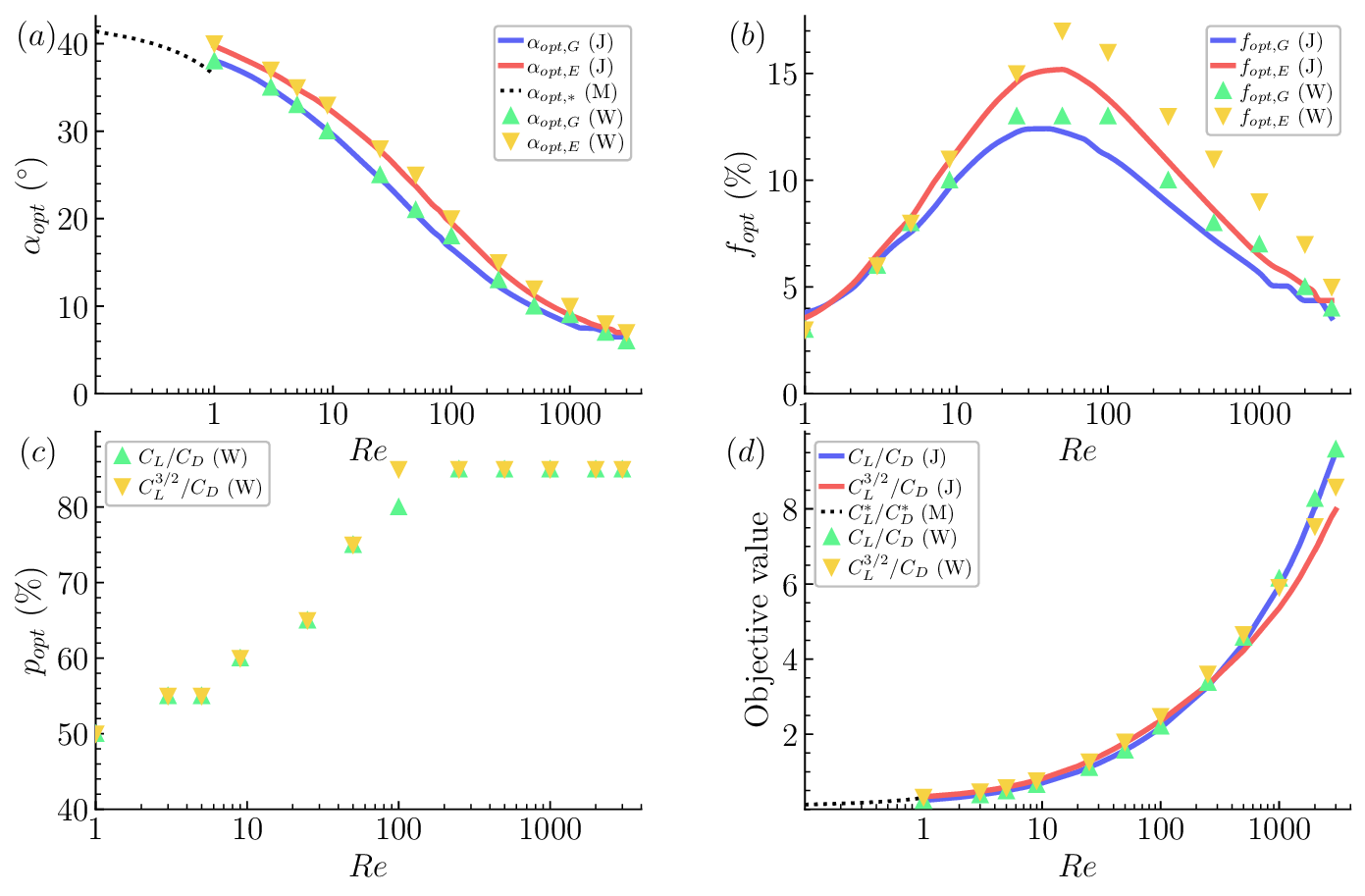}
    \caption{Results of optimization of $C_L/C_D$ (subscript $G$) and
        $C_L^{3/2}/C_D$ (subscript $E$). (a) Optimal angle of attack $\alpha_{opt}$ as a function of $Re$;
        (b) optimal camber $f_{opt}$ as a function of $Re$;
        (c) optimal chordwise position $p_{opt}$ of the maximum camber as a function of $Re$;
        (d) maximum $C_L/C_D$ and $C_L^{3/2}/C_D$
        as functions of $Re$.
        Solid blue and red lines and the letter J correspond to the results obtained for Joukowski airfoils. Green and yellow triangular markers and the letter W correspond to the results of calculations using the Wegmann method with the chordwise position $p$ of the maximum camber treated as one of the optimization parameters.
        Black dotted lines and letter M represent the asymptotic formula for the maximum $C_L^*/C_D^*$ value and corresponding $\alpha_{opt,*}$ for a flat plate.
        % Here J, W, and M denote, respectively, the Joukowski parametrization, the Wegmann parametrization, and the asymptotic flat-plate formula of Miyagi. The Wegmann profile is evaluated at a fixed leading-edge thickness of $1.2\%$ of the chord.}
        The position of the maximum camber of thin Joukowski foils is constrained at $p=50\%$ and therefore not shown on the plot.
    }
    \label{fig:optimalgeometry}
\end{figure}

Let us denote as $\alpha_{opt,G}$ and $f_{opt,G}$, respectively, the angle of attack and camber that maximize $C_L/C_D$.
Similarly, $\alpha_{opt,E}$ and $f_{opt,E}$ will stand for the values that maximize $C_L^{3/2}/C_D$.
We will also use shorter notations  $\alpha_{opt}$ and $f_{opt}$ in statements that apply to both maximization objectives.
The results of the high-resolution grid search optimization are presented in
Fig.~\ref{fig:optimalgeometry}. Solid lines correspond to the Joukowski airfoils. Point markers correspond to the extended optimal shapes obtained with the position of the maximum camber $p$ as an additional optimization parameter, by employing Wegmann transform. For both airfoil families and both objective functions, the results are qualitatively the same and quantitatively
close. Let us first discuss the results obtained for Joukowski airfoils. As $Re$ increases from $1$ to
$3000$, the $\alpha_{opt}$ decreases monotonically from
$39^\circ$ to $6^\circ$, with a maximum difference between $\alpha_{opt,E}$ and $\alpha_{opt,G}$ of $3.2^\circ$
at $Re = 50$. The $f_{opt}$ follows a bell-shape curve, with $f_{opt,G}$ and $f_{opt,E}$ behaving nearly identically at the
boundaries of the $Re$ range considered: starting and ending at $3.5\%$ at $Re = 1$ and $Re = 3000$.
Significant differences appear only near the peaks: $f_{opt,G} = 12.3\%$ at $Re = 40$
and $f_{opt,E} = 15.2\%$ at $Re = 50$, and
a maximum difference of $2.9\%$ at $Re = 50$.

When the position $p$ of the maximum camber is treated as an adjustable parameter, for the flow regimes near $Re = 10$, the optimal position of the maximum camber is near 50\% chordwise. However, for larger $Re$, optimal position of the maximum
camber progressively shifts downstream toward the trailing edge. The cambers $f$ are larger but the angles of attack $\alpha$ are the same as for the Joukowski airfoils. The objective function values are slightly greater than for the Joukowski airfoils. The latter results, however, have been obtained with the fixed maximum thickness near the leading edge of 1.2\% of the chord length as required for the numerical mapping, and this value is considerably larger than the Joukowski airfoil thickness.
It can be postulated that trends shown in Fig.~\ref{fig:optimalgeometry} might be generally representative of optimal thin airfoils.

In the low-Reynolds-number regime, the Oseen approximation is valid, and the numerical
results show that $f_{opt} \to 0$, meaning the optimal shape reduces to a flat plate.
It is therefore natural to compare the numerical results against the analytical solution derived by
\citet{miyagiOseenFlowFlat1964} for the lift and drag of a flat plate under the Oseen
approximation:
$
    C_D^* = \frac{16\pi}{Re} \frac{2S_1 - 1 - \cos 2\alpha}
    {4S_1^2 - 2(1+\cos 2\alpha)}, \quad
    C_L^* = \frac{16\pi}{Re}\frac{\sin 2\alpha}
    {4S_1^2 - 2(1+\cos 2\alpha)},$
where $S_1 = \ln(16/Re) - \gamma$, with $\gamma$ denoting the Euler--Mascheroni constant.
Maximizing $C_L^*/C_D^*$ with respect to $\alpha$ yields the optimal angle,
$\alpha_{opt,*} = \arctan\sqrt{1 - 1/S_1}$,
which satisfies $\alpha_{opt,*} \to 45^\circ$ as $Re \to 0$, explaining the large optimal
angles of attack observed at the lower end of the Reynolds-number range. Since $S_1$ is a
decreasing function of $Re$, $\alpha_{opt,*}$ itself decreases monotonically with $Re$,
consistent with the numerical results.

The non-monotonic behaviour of $f_{opt}$ arises from a competition between positive and negative aerodynamic effects.
The benefits of non-zero camber are twofold. First, curvature enhances the inertial pressure differential between lower and upper surfaces.
Second, these enhanced pressure forces near the leading edge are oriented vertically, contributing to $C_L$ to a greater extent.
On the downside, greater pressure differential on a cambered foil entails larger adverse pressure gradient in the aft portion of it, which leads to flow separation and increased drag.
As $Re$ increases from the lower end, increased inertia boosts the curvature-induced excess $C_L$.
At the same time, viscous diffusion remains strong enough to mitigate flow  separation even at large $f$ and $\alpha$, permitting highly cambered
airfoils to remain near-optimal. Together, these effects drive $f_{opt}$ upward until $Re \approx 40-50$.

\begin{figure}[ht]
    \centering
    \includegraphics[width=0.8\textwidth]{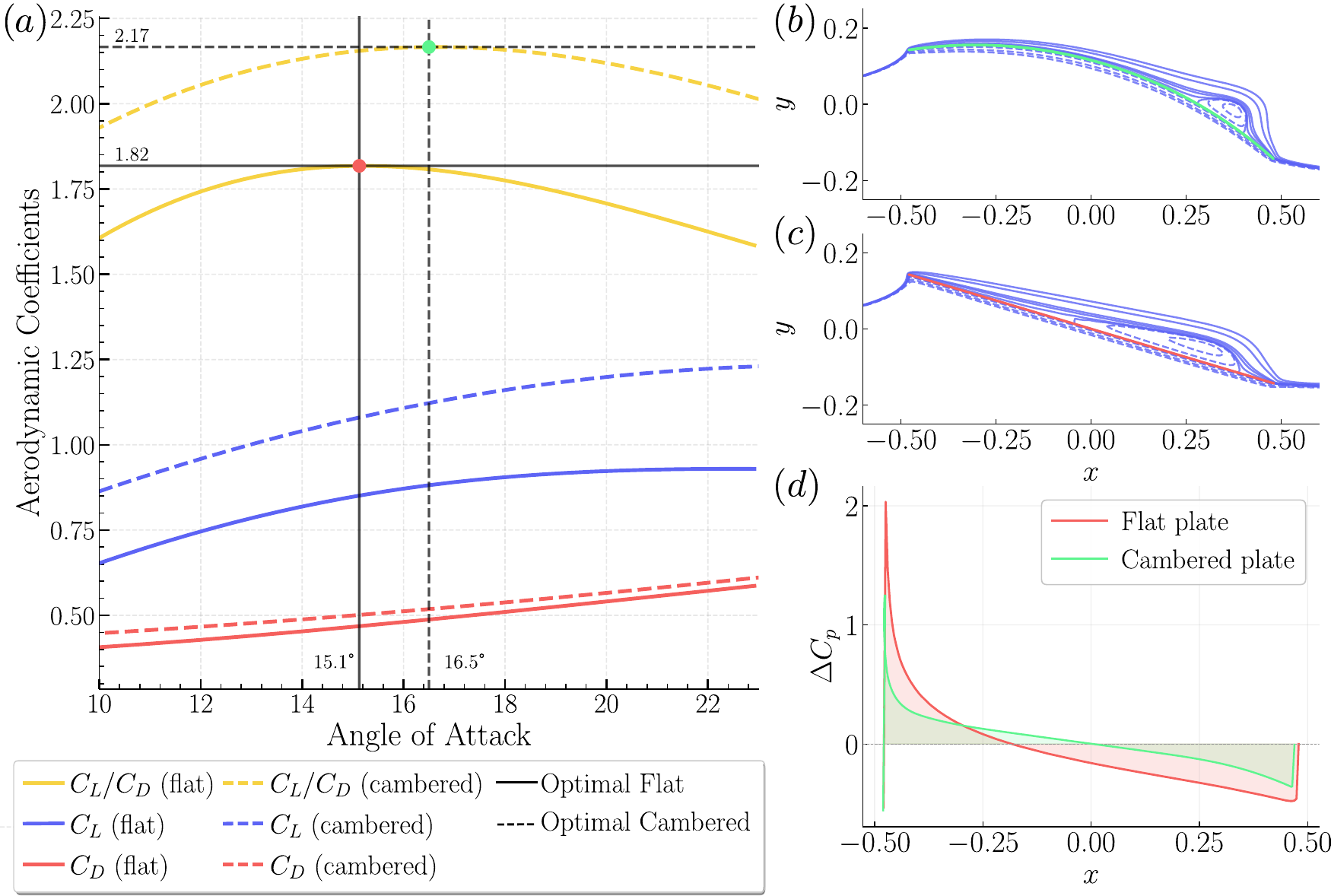}
    \caption{(a)Lift, drag, and lift-to-drag ratio curves as functions of $\alpha$ at $Re = 100$ for the flat and optimal cambered profiles, with optimal $\alpha$ and values of $C_L/C_D$ marked. (b, c)Streamlines for the optimal cambered and flat profiles respectively, shown at levels $0,\pm 10^{-5}, \pm 5\times 10^{-5}, \pm 10^{-4}, \pm 5\times 10^{-4}, \pm 10^{-3}$. (d) Pressure coefficient difference along the chord for both profiles.}
    \label{fig:pressure}
\end{figure}

Beyond this peak, the trend reverses: as inertia becomes dominant, viscosity can no longer resist and the flow separates at large $f$ and $\alpha$ without reattachment.
Both $f_{opt}$ and $\alpha_{opt}$ decline toward lower values.
Flow detachment simultaneously reduces lift and increases drag, impairs
both $C_L/C_D$ and $C_L^{3/2}/C_D$; any $(f, \alpha)$ combination at which the
bubble opens is suboptimal. The optimal airfoils, therefore, reside in the attached-flow regime, which justifies the use of the steady
solver in this study.
At $Re = 100$, the optimal cambered airfoil ($f_{opt,G} = 11.2\%$, $\alpha = 16.6^\circ$ ) already carries
a trailing-edge recirculation bubble, visible in the streamlines
(Fig.~\ref{fig:pressure}b,c), yet the flow remains steady. This airfoil achieves $17\%$ higher $C_L/C_D$ than the flat plate
(Fig.~\ref{fig:pressure}a,d), confirming that moderate camber remains beneficial
even as a separation bubble begins to form.
For the optimal shape found at $Re=3000$ ($f_{opt,G} = 3.5\% $, $\alpha = 6^\circ$), this was confirmed by supplementary unsteady numerical simulations using OpenFOAM (see [\url{https://osf.io/y4pr8}]), in which the fluctuation amplitude did not exceed 0.2\% of the respective mean values.
The Reynolds number based on the perpendicular projection of the chord, $Re_\perp = Re \sin \alpha$, equals 314  in the optimal case at $Re = 3000$, and decreases with the decreasing $Re$, further supporting the steady flow assumption.

\begin{figure}[ht]
    \centering
    \includegraphics[width=\textwidth]{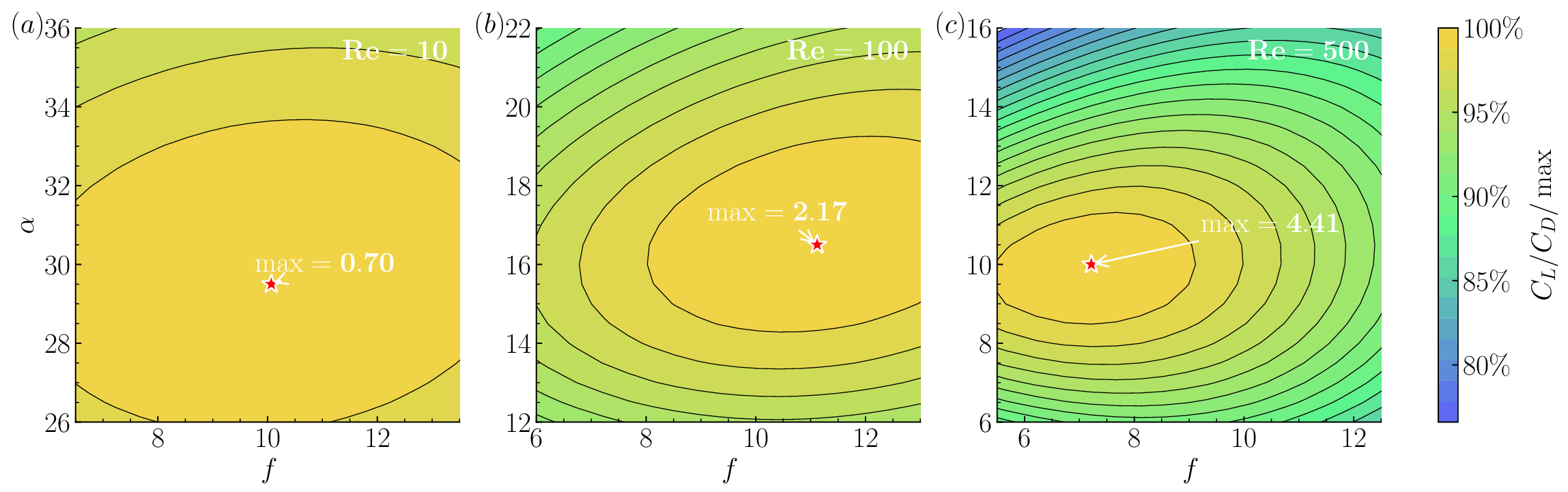}
    \caption{Counterplots of $C_L/C_D$ normalized by its maximum value,
        in the $(f, \alpha)$ parameter space. (a) for $Re = 10$; (b) for $Re = 100$; (c) for $Re = 500$.
        Stars mark the $(f^*, \alpha^*)$ in each case. All panels span
        equal areas of the $(f, \alpha)$ parameter space. }
    \label{fig:optimization_landscape}
\end{figure}

At low $Re$, strong viscous diffusion spreads the
$C_L/C_D$ and $C_L^{3/2}/C_D$ maxima broadly across the $(f, \alpha)$
space; as inertia strengthens with
$Re$, the maxima sharpen, making a geometrically precise optimal airfoil identifiable.
This is illustrated in Fig.~\ref{fig:optimization_landscape}, which shows contour plots of $C_L/C_D$ at $Re = 10, 100$ and $500$. All panels span equal
ranges $\delta f \times \delta\alpha = 7\% \times 10^\circ$, centred on $(f_{opt}, \alpha_{opt})$
at each $Re$, enabling direct comparison of landscape sharpness.
At $Re = 10$, $C_L/C_D$ varies by less than $5\%$ across the displayed range,
meaning that a wide number of $(f, \alpha)$ combinations yields nearly identical
performance. As $Re$ increases, the maximum becomes more distinct: at $Re = 100$
the variation reaches $10\%$, and at $Re = 500$ it is $25\%$.
This progressive sharpening has a direct implication: below $Re \approx 100$,
performance is insensitive to the precise choice of $f$ and $\alpha$, whereas
above $Re \approx 500$, the sharp peak singles out a physically distinguished
$(f, \alpha)$ combination, making optimization both meaningful and effective.

% The flatness of the objective maximum at $Re \lesssim 10$
% (Fig.~\ref{fig:optimization_landscape}a) implies that in this
% regime the performance difference between cambered and flat profiles
% is negligible. It is therefore natural to compare the numerical
% results against the analytical flat-plate solution
% for Oseen flow derived by \citet{miyagiOseenFlowFlat1964}:
% \begin{equation}
%   \label{eq:asymptotic_formula_miyagi}
%   C_D^* = \frac{16\pi}{Re} \frac{2S_1 - 1 - \cos 2\alpha}
%   {4S_1^2 - 2(1+\cos 2\alpha)}, \quad
%   C_L^* = \frac{16\pi}{Re}\frac{\sin 2\alpha}
%   {4S_1^2 - 2(1+\cos 2\alpha)},
% \end{equation}
% where $S_1 = \ln(16/Re) - \gamma$ and $\gamma$ is the Euler constant. The maximum
% of $C_L^*/C_D^*$ and the corresponding optimal angle are shown in
% Fig.~\ref{fig:optimalgeometry}(a,c) for $Re \in [0.1, 1]$ with black dashed lines.
% At $Re = 1$, the asymptotic and numerical optimal angles of attack agree to within
% $1.4^\circ$, and the asymptotic curve extends the trend of maximum $C_L/C_D$
% continuously into $Re < 1$, confirming that the intermediate-$Re$ results connect
% smoothly to the Oseen regime with no anomalous behaviour at the lower boundary of
% the domain.
\section{Conclusion}
Optimization of Joukowski airfoils across $Re \in [1, 3000]$ within the steady, laminar, incompressible Navier–Stokes model
has revealed features and trends of optimal airfoils that are common to both the $C_L/C_D$ and $C_L^{3/2}/C_D$ maximization in the intermediate Reynolds number range.
Zero-thickness cambered airfoils are optimal for both $C_L/C_D$ and
$C_L^{3/2}/C_D$ throughout the entire intermediate-$Re$ range. This result is confirmed by
three independent global search strategies. The optimal angle of attack decreases monotonically with $Re$, while the
optimal camber follows a bell-shaped curve. At $Re \lesssim 10$, near-flat plates
($f \lesssim 11\%$) at high angle of attack ($\alpha \approx 35-39^\circ$)
are optimal. As $Re$ increases, optimal camber attains a peak of $12.3\%$ ($C_L/C_D$)
and $15.2\%$ ($C_L^{3/2}/C_D$) at $Re \approx 40-50$, marking the transition
from viscosity-dominated to inertia-dominated flow. At $Re = 3000$, both $f$ and
$\alpha$ converge toward the slightly cambered (${\sim}4\%$) low-$\alpha$
(${\sim}6^\circ$) geometry, in close agreement with the optimal airfoils identified by
\citet{sunadaAirfoilSectionCharacteristics1997} at $Re = 4000$ and consistent with the slightly cambered
geometry characteristic of conventional low-speed airfoils.
Another useful property is the sharpness of the $C_L/C_D$ and $C_L^{3/2}/C_D$
maxima in $(f,\alpha)$ space. Below $Re \approx 100$, geometric precision is aerodynamically unimportant:
performance varies by less than $10\%$ across a broad neighbourhood of the optimum.
Above $Re \approx 500$, the objective functions have a distinct optimum, with a
$25\%$ change in performance across a neighbourhood of the same size. The optimization then reveals
a distinct optimal shape, clearly superior to neighbouring designs in terms of aerodynamic performance.
After extending the parameter space by including the chordwise position $p$ of the maximum camber as an adjustable parameter, the trends of optimal $\alpha$ and $f$ have remained the same, although the values of $f_{opt}$ have increased by a few per cent.
In addition, these optimization runs have revealed a trend for the optimal chordwise position of the maximum camber to shift, with increasing $Re$, towards the trailing edge, which has not been reported in earlier studies.

\textbf{Declaration of Interests.} The authors report no conflict of interest.

\textbf{AI Statement.} AI tools have been used in the writing process.

\textbf{Data Availability.}
The solver source code is accessible on \href{https://github.com/Zhdanko-Gleb/Optimal-airfoils-in-the-intermediate-Reynolds-number-range}{GitHub}.
Datasets and other supplementary information can be downloaded from [\url{https://osf.io/y4pr8}].
\clearpage
\appendix

\section{Linearization}
\label{app:linearization}

Writing the system~\eqref{eq:TransformedNSwithBC} in operator form as
\begin{equation}
    \label{eq:residual}
    R(\psi, \omega) =
    \begin{pmatrix}
        \dfrac{1}{Re}\nabla^2_{r,s}\omega -
        \left(\dfrac{\partial\psi}{\partial s}\dfrac{\partial\omega}{\partial r}
        - \dfrac{\partial\psi}{\partial r}\dfrac{\partial\omega}{\partial s}\right)
        \\[10pt]
        \dfrac{1}{\det\mathsfbi{J}}\nabla^2_{r,s}\psi + \omega
    \end{pmatrix}
    = 0,
\end{equation}
the Newton update at iteration $n$ seeks a correction $\delta X^n$
satisfying
\begin{equation}
    R'(X^n)[\delta X^n] = -R(X^n), \qquad X^{n+1} = X^n + \alpha^n \delta X^n,
\end{equation}
where $X^n = (\psi^n, \omega^n)$, $\delta X^n = (\delta\psi^n,
    \delta\omega^n)$ is the Newton increment, $\alpha^n \in (0,1]$ is the step
length determined by line search, and $R'(X^n)[\cdot]$ denotes the
Fréchet derivative of the residual operator. The choice $\alpha^n = 1$
recovers the classical Newton step; values smaller than unity correspond to
the damped regime, in which the full step is rejected and a shorter step
along the same direction is accepted instead.

Applying the Fréchet derivative explicitly yields the linearized system
to be solved at each iteration:

\begin{subequations}
    \label{eq:linearisedNS}
    \begin{eqnarray}
        \frac{1}{Re}\nabla^2_{r,s} \delta \omega^n - \left(\pd{\delta \psi^n}{s}\pd{\omega^n}{r}+ \pd{\psi^n}{s}\pd{\delta \omega^n}{r}- \pd{\delta \psi^n}{r}\pd{\omega^n}{s}-\pd{\psi^n}{r}\pd{\delta \omega^n}{s}\right) \\
        \quad = - \frac{1}{Re}\nabla^2_{r,s}\omega^n+\left(\frac{\partial \psi^n}{\partial s}\frac{\partial \omega^n}{\partial r}-\frac{\partial \psi^n}{\partial r}\frac{\partial \omega^n}{\partial s}\right),\\
        \frac{1}{\det \mathsfbi{J}}\nabla^2_{r,s}\delta \psi^n + \delta \omega^n = -\frac{1}{\det \mathsfbi{J}}\nabla^2_{r,s}\psi^n-\omega^n,\\
        \delta \psi^n =  -\psi^n, \text{at }\Gamma \quad (\text{the condition is: } \psi = 0)\\
        \frac{\partial \delta \psi^n}{\partial r} = -\frac{\partial \psi^n}{\partial r}, \text{at }\Gamma \quad (\text{the condition is: } \frac{\partial \psi}{\partial r} = 0)\\
        \pd{\delta \psi^n}{r} = B - \pd{\psi^n}{r}, \text{at }\Gamma_1 \quad (\text{the condition is: } \pd{\psi}{r} = B)\\
        \delta \omega^n = -\omega^n, \text{at }\Gamma_1 \quad(\text{the condition is: } \omega = 0)
    \end{eqnarray}
\end{subequations}

The right-hand sides carry the current residuals of the governing
equations. The boundary conditions on the increments follow directly from
linearizing their nonlinear counterparts: since the boundary values are
fixed, any deviation of $\psi^n$ or $\omega^n$ from the prescribed data
must be corrected in full by the corresponding increment.

\section{Solver verification}\label{app:validation}
\begin{figure}[h]
    \centering
    \includegraphics[width=\textwidth]{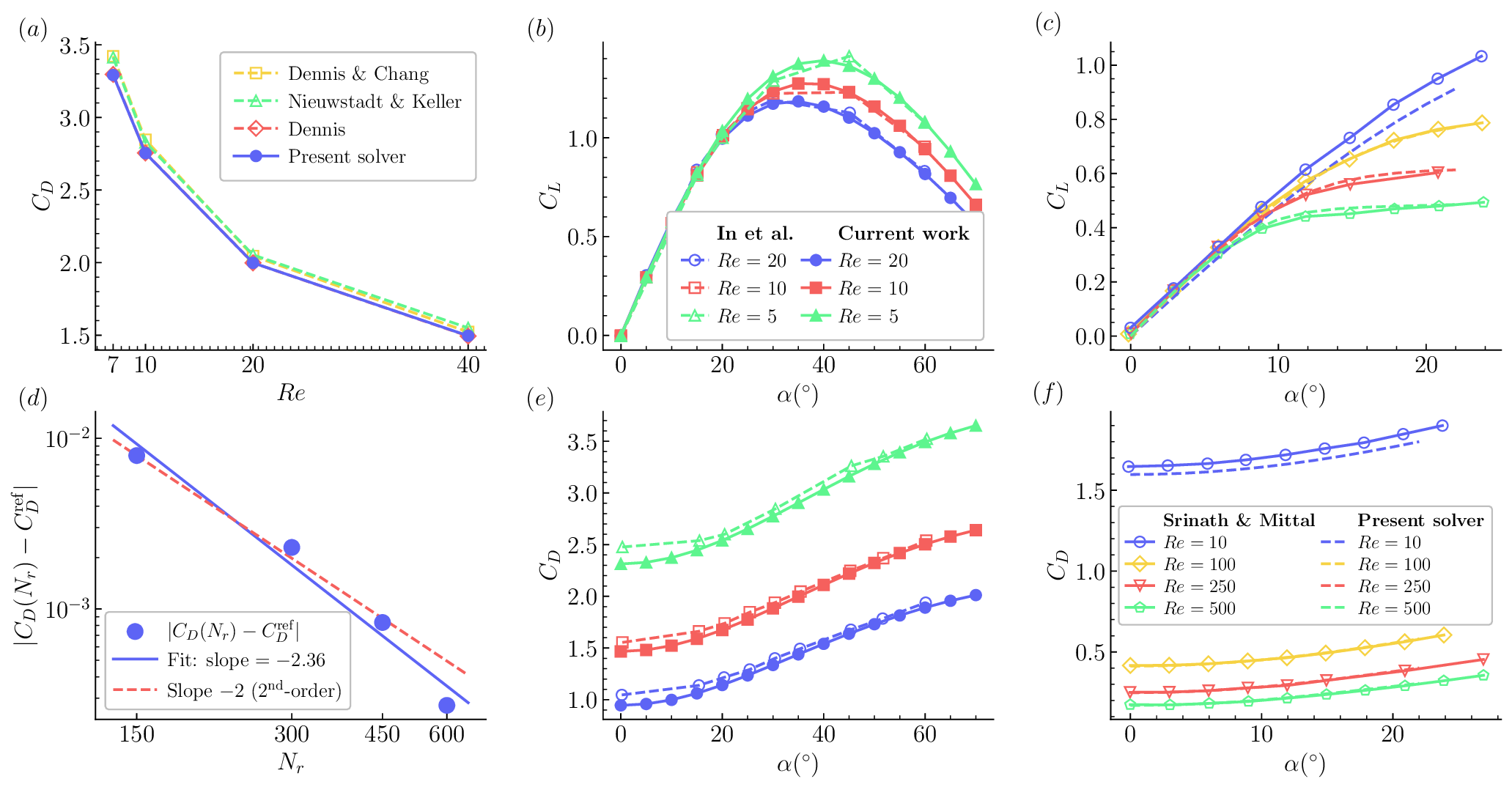}
    \caption{Solver validation. (a) Drag coefficient for flow past a circular cylinder at $Re = 7, 10, 20, 40$, compared with \citet{DennisChang1970}, \citet{NieuwstadtKeller1973}, and \citet{dennisNumericalMethodCalculating1976}. (b, e) Lift and drag coefficients versus angle of attack $\alpha$ for a thin flat plate at $Re = 5, 10, 20$, compared with \citet{inTwodimensionalViscousFlow1995}. (c, f) Lift and drag coefficients versus $\alpha$ for NACA0012 at $Re = 10, 100, 250, 500$, compared with \citet{srinathOptimalAirfoilShapes2009}. (d) Mesh convergence at $Re = 40$, showing second-order behaviour.}
    \label{fig:validation}
\end{figure}

The solver is implemented in Python and is validated against three
benchmark cases: flow past a circular cylinder, flow past a thin flat
plate, and flow past a NACA~0012 airfoil. Figure~\ref{fig:validation}(a)
shows drag coefficients for the cylinder at $Re = 7$--$40$ in close
agreement with the solutions of \citet{DennisChang1970},
\citet{NieuwstadtKeller1973}, and
\citet{dennisNumericalMethodCalculating1976}.
Figures~\ref{fig:validation}(b, e) compare lift and drag coefficients for
the flat plate at $Re = 5, 10, 20$ with results of
\citet{inTwodimensionalViscousFlow1995}, showing good agreement across
all $\alpha$. Figures~\ref{fig:validation}(c, f) compare lift and drag
coefficients for the NACA~0012 airfoil at $Re = 10, 100, 250, 500$ with
results of \citet{srinathOptimalAirfoilShapes2009}. The coefficients
agree closely at $Re = 100, 250, 500$ for all $\alpha$; at $Re = 10$
the agreement is weaker, which we attribute to the comparatively small
computational domain (10 chord lengths) used by
\citet{srinathOptimalAirfoilShapes2009}, insufficient to fully capture
the far-field decay of the flow at this low Reynolds number. Mesh
convergence is demonstrated in Figure~\ref{fig:validation}(d) for the
cylinder at $Re = 40$, confirming second-order spatial accuracy.

\clearpage
\bibliography{references}

\end{document}